\pdfoutput=1
\documentclass[sigconf]{acmart}
\usepackage{colortbl}
\usepackage[colorinlistoftodos]{todonotes}
\usepackage{comment}
\usepackage{algorithm}
\usepackage{algorithmic}
\usepackage{pgfplots}
\usepackage{xcolor}
\usepackage{array}
\newcolumntype{g}{>{\columncolor{lightgray}}c}

\AtBeginDocument{%
  \providecommand\BibTeX{{%
    \normalfont B\kern-0.5em{\scshape i\kern-0.25em b}\kern-0.8em\TeX}}}

\setcopyright{acmcopyright}
\copyrightyear{2020}
\acmYear{2020}
\acmDOI{}

\acmConference[aaa '20]{aa '20: aa}{2020}{USA}
\acmBooktitle{aaa '20: aa,
  June 03--05, 2020, USA}
\acmPrice{15.00}
\acmISBN{}

\begin{document}

\acmYear{2020}
\setcopyright{acmcopyright}\acmConference[aa '20]{aa 2020}{September 29-October 2, 2020}{Virtual Event, CA, USA}
\acmBooktitle{aa 2020 (aa '20), September 29-October 2, 2020, Virtual Event, CA, USA}
\acmPrice{15.00}
\acmDOI{10.1145/3395027.3419589}
\acmISBN{978-1-4503-8000-3/20/09}

\title{Fast Clustering of Short Text Streams Using Efficient Cluster Indexing and Dynamic Similarity Thresholds}

\author{Md Rashadul Hasan Rakib}
\affiliation{%
  \institution{Dalhousie University}
  \city{Nova Scotia}
  \country{Canada}}
\email{rakib@cs.dal.ca}

\author{Muhammad Asaduzzaman}
\affiliation{%
  \institution{Queen's University}
  \city{Ontario}
  \country{Canada}}
\email{muhammad.asaduzzaman@cs.queensu.ca}


\renewcommand{\shortauthors}{rakib et al.}

\begin{abstract}
Short text stream clustering is an important but challenging task since massive amount of text is generated from different sources such as micro-blogging, question-answering, and social news aggregation websites.
One of the major challenges of clustering such massive amount of text is to cluster them within a reasonable amount of time. 
The existing state-of-the-art short text stream clustering methods can not cluster such massive amount of text within a reasonable amount of time as they compute similarities between a text and all the existing clusters to assign that text to a cluster. 
To overcome this challenge, we propose a fast short text stream clustering method (called FastStream) that efficiently index the clusters using inverted index and compute similarity between a text and a selected number of clusters while assigning a text to a cluster.
In this way, we not only reduce the running time of our proposed method but also reduce the running time of several state-of-the-art short text stream clustering methods.
FastStream assigns a text to a cluster (new or existing) using the dynamically computed similarity thresholds based on statistical measure.
Thus our method efficiently deals with the concept drift problem.
Experimental results demonstrate that FastStream outperforms the state-of-the-art short text stream clustering methods by a significant margin on several short text datasets. 
In addition, the running time of FastStream is several orders of magnitude faster than that of the state-of-the-art methods.

\end{abstract}





\maketitle

\section{Introduction}
\label{sec:introduction}
Due to technological advances short text streams are generated at large volumes from different sources. 
Organizing these text streams within reasonable time-frame is an important step towards discovering real-time trends (e.g., political, economic) in conversations, finding groups of users sharing similar topics of interest~\citep{usercluster:2016}, and monitoring the activity of individuals over time. 
The task of clustering short text streams is to assign a text to a new cluster or to one of the existing clusters as it arrives~\citep{Mstream:2018}. 
The recent short text stream clustering methods were proposed based on dirichlet process multinational mixture model as described in~\citep{Mstream:2018, osdm:2020, BMM2020}. 
Generally, these algorithms used Gibbs Sampling~\citep{gibssampling:2001} to estimate the parameters of the mixture model so as to obtain the clustering of text streams.
These algorithms assign a text to a cluster by computing similarities between a text and the clusters based on the common features between them (e.g., word, biterm).
However, there may be some clusters with which a text may not share any common feature, thus the similarities between that text and those clusters are zero.
Therefore, those similarity computations can be ignored.


Motivated by this, FastStream selects a specific set of clusters while computing similarities between a text and the clusters. 
These selected clusters share common features with a particular text.
By limiting the number of similarity computations, we significantly reduce the running time of our proposed method and thus the running time of our method is several orders of magnitude faster than that of the state-of-the-art methods.

In general, similarity-based text stream clustering methods use user defined similarity threshold to assign a text to a new or one of the existing clusters~\citep{Mstream:2018}.
\citet{rakib:2020} proposed a short text stream clustering method that removes outliers from the clusters and reassigns them to the clusters using dynamically computed similarity thresholds~\citep{rakib:2020}.
This motivates us to dynamically calculate the similarity threshold for each text based on statistical measure and use this similarity threshold to assign the text to a new or one of the existing clusters.
Thus our method efficiently handles the \emph{concept drift} problem~\citep{Mstream:2018} (the problem that topics of the text streams may change over time).





The three major contributions of our work are as follows.
\begin{itemize}

\item First, we improve the running time of our method\footnote{https://github.com/rashadulrakib/short-text-stream-clustering/tree/master/OnlineClustering} by computing the similarity between a text and a selected number of clusters (obtained by inverted index~\citep{invertedindex:2014}) instead of all clusters. Therefore, the running time of our method is several orders of magnitude faster than that of the state-of-the-art short text stream clustering methods.
Moreover, the experimental results demonstrate that by applying inverted index, we are able to reduce the running time of several state-of-the-art methods.


\item Second, our method utilizes dynamically calculated similarity thresholds based on the statistical measure to assign a text to a new or an existing cluster. The use of dynamic thresholds helps to avoid the concept drift problem~\citep{Mstream:2018}. 

\item Third, we perform an extensive comparison of our proposed method using four different datasets and our method outperforms the state-of-the-art short text stream clustering methods by a significant margin on several datasets.
\end{itemize}


\section{Related Work}
\label{sec:relatedwork}

\subsection{Similarity-based Stream Clustering}
\label{subsec:relatedwork_similarity}

In general, similarity-based text stream clustering methods use the vector space model~\citep{VSM2012} to represent the documents.
A document is assigned to a new or one of the existing clusters based on the similarity threshold which needs to be manually determined by the user~\citep{Mstream:2018}. 
A detailed survey of clustering data streams can be found in~\cite{DataSurvey2019}.
In our method, we dynamically calculate the similarity threshold for each text and use this similarity threshold to assign the text to a new or one of the existing clusters.

CluStream~\citep{clustream:2003} is a stream clustering method consisting of an online micro-clustering phase and an offline macro-clustering phase. 
In online phase, it assigns a data point to a new or an existing micro-cluster based on a similarity threshold. 
In offline phase, it applies k-means clustering on the micro-clusters and obtain k user specified macro-clusters.

Sumblr~\citep{Sumblr:2013} is a tweet stream summarization prototype. 
It consists of a text stream clustering module that compresses tweets into tweet feature vectors (TCVs) and assigns future tweets to the clusters based on the statistics of TCVs.

An efficient text stream clustering method using term burst information was proposed in~\citep{textstream2016}.
Bursty terms are the terms that appear in many documents during a short period of time.
This approach considers the fact that the documents that are published on a particular topic within a certain time period contain a particular set of bursty terms.
An user-defined threshold for the number of occurrences of terms in documents is used to identify bursty terms.

\subsection{Model-based Stream Clustering}
\label{subsec:relatedwork_model}

A recent short text stream clustering algorithm based on dirichlet process multinational mixture model was proposed in \citep{Mstream:2018} which uses two dirichlet priors $\alpha$ and $\beta$. 
$\alpha$ refers to the prior probability of a text choosing a new cluster and $\beta$ corresponds to the prior probability of a text choosing a cluster with which the text shares more similar content than other clusters.
This algorithm has two variants: one is by retaining all previous clusters (called MStream) and other one is by removing old clusters (called MStreamF).

\citet{rakib:2020} proposed a short text stream clustering method that clusters a fraction of texts in each batch\footnote{1A Batch (or stream) is defined as a collection of texts~\citep{Mstream:2018}. The terms ``Batch'' and ``Stream'' are interchangeably used in our paper} using the frequent biterms in texts, then it populates the clustering model of MStream algorithm using the cluster assignments obtained by the frequently occurred biterms in texts.
After that it clusters rest of the texts in the batch using the populated clustering model of MStream algorithm.
Following that it removes outliers from the clusters and reassigns them to the clusters (new or existing) using the dynamically computed similarity thresholds.

A biterm based mixture model for short text stream clustering was proposed in~\citep{BMM2020}. 
Similar to MStream(F) algorithm~\citep{Mstream:2018}, the biterm based clustering method developed two variants: one is by retaining the clusters obtained in previous batches (called DP-BMM) and other is by discarding the clusters obtained in previous batches (called DP-BMM-FP). 
The main difference between MStream(F) and DP-BMM(-FP) is that DP-BMM(-FP) represents the texts using biterm features instead of unigrams. 
In particular, DP-BMM(-FP) represents a text of $n$ words using $n*(n-1)/2$ biterm features.

OSDM~\citep{osdm:2020} is a semantic-enhanced dirichlet model for short text stream clustering. 
OSDM extends the MStream~\citep{Mstream:2018} algorithm by integrating the word to word co-occurrence based semantic information obtained from the common words between a text and a cluster and uses this semantic information to compute similarity between a text and a cluster.

In general, a Dirichlet process mixture model based clustering algorithm requires tuning the parameters (i.e., $\alpha$ and $\beta$) to obtain the desired clustering performance. 
For example, MStream(F) uses $\alpha=0.03$ and $\beta=0.03$ to obtain optimal clustering performance. 
The DP-BMM(-FP) and OSDM performed grid search to obtain the optimal values for $\alpha$ and $\beta$ and set ($\alpha=0.6$ and $\beta=0.02$) and ($\alpha=2e^{-3}$ and $\beta=4e^{-5}$) respectively.
On the contrary, our proposed method (FastStream) does not require this kind of parameter tuning, instead it uses the dynamically computed similarity threshold (based on statistical measure) to assign a text to a new or an existing cluster.

DCT-L~\citep{DCT-L:2016} is a dynamic clustering topic model for short text streams based on dirichlet process multinomial mixture model.
It assigns a single topic (i.e., cluster) to each short text at a particular timestamp and uses the resulting topic distribution as priors for inferring the topics of subsequent documents.

DTM~\citep{DTM:2006} is a dynamic topic model that analyzes the topics of a collection of documents over time. 
This method assumes that a document is rich enough to contain multiple topics. 
However, this assumption does not work well for short texts, which results low quality performance on short text streams.

\section{Background}
\label{sec:background}
\subsection{Text Representation}
\label{subsec:textrep}
We represent each text using three different kinds of features ($f$) which are unigram, bigram and biterm and use each representation separately to cluster the streams of texts.
The words of a text are considered as the unigram features of that text.
The two consecutive words of a text are considered as the bigram features of that text.
The biterm feature is defined as an unordered \emph{word pair} constructed using the words contained in a text~\citep{BMM2020}.
For example, the text ``ai improves healthcare system'' will be represented by the following bigrams: ``ai improves'', ``improves healthcare'', and ``healthcare system''. 
The same text will be represented by the following biterms: ``ai improves'', ``ai healthcare'', ``ai system'', ``improves healthcare'', ``improves system'', and ``healthcare system''. 

\subsection{Cluster Representation}
\label{subsec:clusrep}

In our method, each cluster is represented by a cluster feature ($CF$) vector~\citep{Mstream:2018} consisting of 4 tuples \{$n_{z}^f$, $n_{z}$, $m_{z}$, $id_{z}$ \} where $n_{z}^f$ refers to the features (unigram, bigram or biterm) along with frequencies in cluster $z$, $n_{z}$ refers to the number of features in cluster $z$, $m_{z}$ refers to the number of texts in cluster $z$, and $id_{z}$ refers to the unique \emph{id} for cluster $z$.






\subsection{Inverted Index}
\label{subsec:invertindex}
Inverted Index is a hashmap like data structure that creates mapping from document-features (unigram, bigram or biterm) to documents~\citep{invertedindex:2014}.
To keep track of which clusters are associated with which features, we adopt the inverted index based searching technique and create a Feature vector $F$ for each feature, defined as a tuple of \{$l_{f}^{id}$\} where $l_{f}^{id}$ refers to the list of cluster $ids$ associated with a feature $f$. 

\section{Proposed Method}
\label{sec:ProposedAlgorithm}

The proposed method (FastStream) clusters each short text one by one as it arrives. 
At first, the features are extracted from the text.
At a time, we use only one type of feature to cluster the streams of texts.
That means, we extract only either unigram, bigram or biterm features from the text.
After that we select the clusters that contain the features of the text.
Then our method computes similarities between the text and the selected clusters using common features.
Following that it assigns the text to an appropriate cluster (new or existing) using the dynamically computed similarity thresholds based on statistical measure.
After that it builds a clustering model using the cluster assignment of the text to reflect the addition of this text to a new or an existing cluster; and uses the current clustering model to cluster the subsequent text.
The details are shown in Algorithm~\ref{algo:shorttextstreamclus} and are described next.
\begin{algorithm}
\footnotesize
\caption{Proposed Text Stream Clustering}
\label{algo:shorttextstreamclus}
\hspace*{\algorithmicindent} \textbf{Input:} Texts: $t_1 ... t_\infty$, $DI$: Delete-interval\\
\hspace*{\algorithmicindent} \textbf{Output:} Cluster assignments: $z_{t_1 ... t_\infty}$ 

\begin{algorithmic}[1]
\FOR {$t_i$ in $t_1 ... t_\infty$}
	\STATE Extract features ($f$) from $t_i$ 
	\STATE Select $L$ clusters that share common features with $t_i$ (described in Section~\ref{sec:selecttargetcluster})
    
	\STATE Compute similarities ($s_{l}$) between $t_i$ and the $L$ selected clusters
	
	\STATE Compute the maximum ($\max_l$), mean ($\mu_l$) and standard deviation ($\sigma_l$) of the $s_{l}$ similarities
	
\IF{$\max_l > \mu_l + \sigma_l $}

\STATE $j=\,$ cluster index for $\max_l$
\STATE Assign $t_i$ to $j^{th}$ cluster
\ELSE
\STATE Assign $t_i$ to a new cluster
\ENDIF
\STATE Build clustering model (described in Section~\ref{sec:buildclusteringmodel})
\IF{$i\mod DI=0$} 
\STATE Delete outdated clusters (described in Section~\ref{sec:deleteoutdatedclus})
\ENDIF
\ENDFOR
\end{algorithmic}
\label{algo-sparsify}
\end{algorithm}

\subsection{Selecting Clusters for the Text}
\label{sec:selecttargetcluster}
For each short text, we select a specific set of clusters that share common features with the text.
For each feature in a text, we obtain the cluster $ids$ from the corresponding feature vector $F$.
We combine the cluster $ids$ obtained using the features in a text.
These are the selected clusters that share common features with the text. 

\subsection{Computing Similarities between the Text and Selected Clusters}
\label{sec:computesimbtntextclusters}
FastStream computes similarities between the text and the selected clusters based on the common features as shown in Equation~\ref{eq:similarity}. 
\begin{equation} 
\label{eq:similarity}
 similarity(t, z)=\frac{2\times\sum_{f \in t}\min (n_{z}^f,N_{t}^f)}{N_t+n_z}  
\end{equation}
To compute similarity between a text $t$ and a cluster $z$, we sum the occurrences of the common features between $t$ and $z$ which is then normalized by the summation of the total number of features in text $t$ and cluster $z$ denoted by $N_t$ and $n_z$ respectively. 
Here $n_{z}^f$, $N_{t}^f$ refer to the features ($f$) along with frequencies in cluster $z$ and text $t$ respectively.

\subsection{Assigning Text to Cluster}
\label{sec:assigntexttocluster}
To assign a text to a cluster, we compute the maximum ($max$), mean ($\mu$) and standard deviation ($\sigma$) of the similarities between the text $t$ and the selected clusters.
We assign the text to the cluster with the maximum similarity if the maximum similarity is greater than the $\mu+\sigma$ of the similarities.
Otherwise, we create a new cluster containing this text.
Thus the texts are assigned to clusters based on the dynamically computed similarity thresholds.
The intuition behind using maximum similarity greater than $\mu+\sigma$ is that, this maximum similarity is above the average similarities reflecting that both the text and the target cluster share highly similar content.

\subsection{Building Clustering Model}
\label{sec:buildclusteringmodel}
We build clustering model using the cluster assignment of the text to reflect the addition of this text to an existing or a new cluster.
When a text $t$ is added to a cluster $z$, we update the corresponding $CF$ vector by updating its features with frequencies ($n_{z}^f$), number of features ($n_z$), and number of texts ($m_z$).
The addible property of the $CF$ vector is described in the following. 

\paragraph{Addible Property of Text to Cluster:}
$\\
n_{z}^f=n_{z}^f+N_{t}^f \;\;\; 
\forall f\in\mathbb \,t
\\ n_{z}=n_{z}+N_{t}
\\ m_{z}=m_{z}+1
\\
id_{z}=
    \begin{cases}
      id_{z}, \; \text{if successive texts are in same cluster} \\
      max(id_{z})+1 \;\;\; 
\forall z\in\mathbb \,Z, \; \text{otherwise}
    \end{cases} \\ \\
$

Here, $N_{t}^f$ and $N_{t}$ refer to the features with frequencies in text $t$ and the total number of features in text $t$ respectively.
$Z$ is the set of all $CF$ vectors (i.e., $Z=\{CF\}$).

A new $id$ is assigned to the cluster $z$ (new or existing) if the cluster assignment of the current text is different from that of the previous text, otherwise the cluster $id$ of the current text remains same as that of the previous text. 
This implies that the recently created or updated cluster will have the highest cluster $id$.
For each feature in the text, we append the cluster $id$ to the corresponding feature vector $F$.

\subsection{Deleting Outdated Clusters}
\label{sec:deleteoutdatedclus}

We remove the outdated clusters based on their update-timestamps (represented by cluster $id$) and cluster-sizes (i.e., number of texts in clusters).
The recently created or updated clusters will have higher cluster $ids$.

We remove outdated clusters in every Delete-interval.
Delete-interval is the interval when we remove outdated clusters after clustering a certain number of texts.
For example, the Delete-interval equal to 500 implies that we delete outdated clusters after we cluster every 500 texts. 
To obtain outdated clusters, we calculate the $\mu$ and $\sigma$ of the cluster $ids$ and cluster-sizes in every Delete-interval.
If the \emph{cluster id} is less than the $\mu - \sigma$ of cluster $ids$ and the cluster-size is less than the $\mu - \sigma$ of cluster-sizes, then we delete the cluster by deleting the corresponding $CF$ vector and remove the corresponding cluster $id$ from the feature vectors ($F$) that contain that particular cluster $id$.

\section{Experimental Study}
\label{sec:experimentalstudy}

\subsection{Datasets}
We used four different datasets of short texts in our experiments. The basic properties of these datasets are shown in Table~\ref{tbl_data_summary}.
\begin{table}[H]
\centering
\scriptsize
\captionsetup{font=small}
\caption{Summary of the short text datasets}
\label{tbl_data_summary}
\renewcommand{\tabcolsep}{2pt}
\begin{tabular}{|c|c|c|c|}
    \hline
Dataset & \#Clusters & \#Texts & Avg. \#words/text\\
\hline
Ns-T & 152 & 11,109 & 6.23\\ 
\hline
Ts-T & 269 & 30,322 & 7.97\\ 
\hline
SO-T & 10,573 & 1,23,342 & 5.57\\ 
\hline
NTSO-T & 10,994 & 1,64,773 & 6.02\\
\hline
\end{tabular}
\end{table}
The datasets \textbf{Ns-T}~\cite{Mstream:2018} and \textbf{Ts-T}~\cite{Mstream:2018} consist of 11,109 news titles and 30,322 tweets and are distributed into 152 and 269 groups respectively.
We create a new dataset called \textbf{SO-T} using the titles of the StackoverFlow questions consisting of 1,23,342 question titles distributed into 10,573 clusters.
We combined the texts of the datasets \textbf{Ns-T}, \textbf{Ts-T}, and \textbf{SO-T} and created a combined dataset \textbf{NTSO-T} consisting of 1,64,773 texts distributed into 10,994 clusters.
The average number of words per text in the datasets \textbf{Ns-T}, \textbf{Ts-T}, \textbf{SO-T}, and \textbf{NTSO-T} are 6.23, 7.97, 5.57, and 6.02 respectively.
The texts in these four datasets were randomly shuffled to examine how FastStream and the state-of-the-art methods  (MStream(F)~\cite{Mstream:2018}, DP-BMM(-FP)~\cite{BMM2020}, and OSDM~\cite{osdm:2020}) perform when dealing with the texts from different domains arriving in random order.

\subsubsection{Construction of the Dataset SO-T}
\label{sub_so_dataset_summary}
We create a dataset \textbf{SO-T} using the titles of the duplicate questions posted in StackOverflow website\footnote{https://stackoverflow.com/} on various topics such as \emph{Java}, \emph{Python}, \emph{JQuery}, \emph{R}, and so on.
We consider that duplicate questions are similar to each other and a group of similar questions can form a cluster.

We obtain the question titles from the file Posts.xml\footnote{https://meta.stackexchange.com/questions/2677/database-schema-documentation-for-the-public-data-dump-and-sede} and obtain the information about the duplicate questions from PostLinks.xml.
Each item in Posts.xml represents a single post which can be of different types (e.g.,  question, answer, and so on).
Each item of the PostLinks.xml contains the information about a pair of duplicate questions.
For instance, the PostLinks.xml contains the questions A and B if they are duplicate.
There are 20,094,655 questions in Posts.xml and 1,009,249 pair of duplicate questions in PostLinks.xml.
Among the 1,009,249 pair of duplicate questions, we randomly select 400,000 pairs\footnote{We select this specific number of pairs (400,000) because of the maximum capacity of the computer (Core i5-4200U and 8GB memory) where the experiments were carried out.}.

Using the duplicate information in PostLinks.xml, we create a list of directed edges (e.g., $A\rightarrow B$, $B\rightarrow C$) which are then used to create a graph.
To obtain the clusters of duplicate questions, we compute connected components\footnote{We use the library in https://networkx.github.io/ to compute connected components.} using the representation of the graph.
In particular, If A and B are duplicate, and B and C are duplicate, then we obtain the connected component $A\rightarrow B\rightarrow C$ which is considered as a cluster of the duplicate questions of A, B, and C.

We compute the length of the connected components defined as the number of questions in the component. 
After that, we compute the mean ($\mu$) and standard deviation ($\sigma$) of the lengths of the connected components and select the components whose lengths are between the $\mu\pm\sigma$ which in turn produces 10,573 connected components (i.e., clusters) consisting of 1,23,342 question titles.
Sample StackOverflow question titles (with \emph{PostId}) of a cluster in the dataset \textbf{SO-T} are shown in Figure~\ref{fig:samplequestion}.
\begin{figure}[H]

\fbox{\begin{minipage}{20em}
-- \footnotesize Python Video Framework (1003376) \\
-- \footnotesize Best video manipulation library for Python? (220866) \\
-- \footnotesize Trim (remove frames from) a video using Python (7291653)
\end{minipage}}
\captionsetup{font=small}
\caption{Sample StackOverflow question titles  (with \emph{PostId}) of a cluster in the dataset \textbf{SO-T}.} \label{fig:samplequestion}
\end{figure}

\subsection{Proposed Clustering Method with Different Types of Features}
\label{sub:propclusdifffeature}
We represent each text using three different types of features which are unigram, bigram and biterm and use each representation separately to cluster the streams of texts.
When we use unigram, bigram or biterm feature, our method is called as \textbf{FastStream-unigram}, \textbf{FastStream-bigram}, and \textbf{FastStream-biterm} respectively.

\subsection{Optimal Delete-interval for Proposed Method}\label{sub:deleteinterval}
Our proposed method (FastStream) requires only one parameter called Delete-interval ($DI$) to delete the outdated clusters.
We set Delete-interval to 500 for all the datasets used in our experiments.
How we choose this value is discussed in the following.

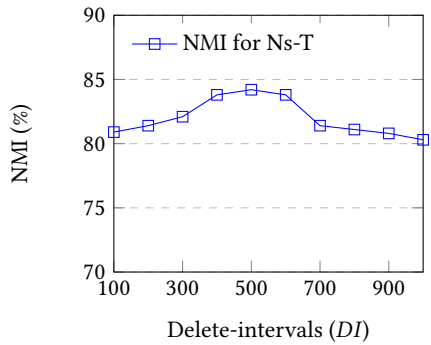
\begin{figure}
\begin{tikzpicture} 
\begin{axis}[
    title={},
    xlabel={Delete-intervals ($DI$)},
    ylabel={NMI (\%)},
    xmin=100, xmax=1000,
    ymin=70, ymax=90,
    xtick={100,300,500,700,900,1100},
    ytick={70, 75, 80, 85, 90},
    legend pos=north west,
    ymajorgrids=true,
    grid style=dashed,
    legend style={draw=none},
    scale = .60
]
 
\addplot[
    color=blue,
    mark=square,
    ]
    coordinates {
    (100,80.9)(200,81.4)(300,82.1)(400,83.8)(500,84.2)(600,83.8)(700,81.4)(800,81.1)(900,80.8)(1000,80.3)
    };
     \addlegendentry{NMI for Ns-T}

\end{axis}
\end{tikzpicture}

\captionsetup{font=small}
\caption{NMI results of our short text stream clustering method for different Delete-intervals ($DI$) on dataset Ns-T.} \label{figtunedeleteinterval}
\end{figure}

Delete-interval is the interval when we remove outdated clusters after clustering a certain number of texts.
To determine the optimal value of Delete-interval, we choose the dataset Ns-T and represent the text using biterm features.
We determine the value of Delete-interval based on the optimal clustering performance of our method for the dataset Ns-T; and use this value to delete outdated clusters for other datasets.
The clustering performance (in terms of NMI) of our method for different Delete-intervals on the dataset Ns-T is shown in Figure~\ref{figtunedeleteinterval}. 

Based on the different values of $DI$, we observe that we achieve the highest NMI for Ns-T when $DI$=500.
Therefore, we choose $DI$=500 for all datasets used in our experiments to delete outdated clusters.

\subsection{Baseline Clustering Methods}
\label{subsec:baselineclusmethods}
We compare the performance of FastStream with recent state-of-the-art short text stream clustering methods as described in the following.

\begin{itemize}
    \item \textbf{MStream}~\citep{Mstream:2018} algorithm clusters each batch of short texts at a time.
    It stores all the clusters produced over the course of time.
    MStream has one pass clustering process and update clustering process of each batch.
    In update clustering process, it applies gibbs sampling to the same batch of texts multiple times to improve the initial clustering result obtained in one pass clustering process.
    
    \item \textbf{MStreamF}~\citep{Mstream:2018} is a variant of MStream algorithm that deletes the outdated clusters of previous batches and only stores the clusters of the current batch.
   
    \item \textbf{DP-BMM}~\citep{BMM2020} is a short text stream clustering algorithm that adopts the similar approach as MStream and clusters each batch of short texts at a time as it arrives.
    The principal difference between DP-BMM and MStream is that DP-BMM represents the texts using biterm features instead of unigrams (i.e., words). 
    Therefore, DP-BMM stores the clusters of texts using biterm features.
    
    \item \textbf{DP-BMM-FP}~\citep{BMM2020} is a variant of DP-BMM  algorithm that deletes the outdated clusters of previous batches and only stores the clusters of the current batch similar to MStreamF.

    \item \textbf{\citet{rakib:2020}} proposed a short text stream clustering method by adopting the clustering model of MStream algorithm~\citep{Mstream:2018}. The main difference between this method and MStream algorithm is that this method removes outliers from the clusters obtained by MStream algorithm and reassigns the outliers to the clusters using dynamically computed similarity thresholds.
    
    \item \textbf{OSDM}~\citep{osdm:2020} is a short text stream clustering algorithm that clusters each short text one by one as it arrives. 
    OSDM deletes a cluster if a cluster becomes outdated, that is, the cluster is not being updated for a while over time.
    
\end{itemize}

For MStream(F), we set the parameters $\alpha=0.03$ and $\beta=0.03$ for all the datasets as defined in~\citep{Mstream:2018}.
Likewise, for DP-BMM(-FP), we set $\alpha=0.6$ and $\beta=0.02$ as mentioned in~\citep{BMM2020}.
MStream(F) and DP-BMM(-FP) cluster each batch of texts at a time.
We set the batch size\footnote{The batch size equal to 2000 was chosen for MStream(F) and DP-BMM(-FP) based on their optimal performance on the datasets used in this paper.} to 2000 for MStream(F) and DP-BMM(-FP) for all datasets.
We set the number of iterations to 10 for MStream(F) and DP-BMM(-FP), and number of saved batches to one for MStreamF and DP-BMM-FP as mentioned in~\citep{BMM2020}.

For OSDM, we set $\alpha=2e^{-3}$, $\beta=4e^{-5}$, and an additional parameter $\lambda=6e^{-6}$ for all the datasets as defined in~\citep{osdm:2020}.
OSDM used $\lambda$ as the decay rate which is used to set lower weights to the clusters which are not being updated over time and need to be deleted in the future.

\subsection{Faster Version of Baseline Clustering Methods}
\label{subsec:baselineclusmethods}

\begin{itemize}

\item \textbf{Fast-MStream}\footnote{https://github.com/rashadulrakib/short-text-stream-clustering/tree/master/FastBatchClustering} and \textbf{Fast-MStreamF} are the faster versions of MStream and MStreamF algorithms that we developed respectively.
    We apply the inverted index~\citep{invertedindex:2014} based searching technique using the words of the clusters produced by the MStream and MStreamF algorithms.

    
 \item \textbf{Fast-\citet{rakib:2020}} is the faster version of \citet{rakib:2020}.
 We index the clusters produced by \citet{rakib:2020} using unigrams following the notion of Fast-MStream.

\end{itemize}

\subsection{Comparison with State-of-the-art Methods}
\subsubsection{Comparison of Clustering Results}
We compare the performance of our proposed method (FastStream) and the proposed faster versions of the state-of-the-art short text stream clustering methods 
with the state-of-the-art methods.
We apply different types of text representations (unigram, bigram, and biterm) to FastStream denoted as FastStream-unigram, FastStream-bigram, and FastStream-biterm.
We use normalized mutual information (NMI) as the evaluation measures for evaluating the performance of different clustering methods.
We randomly shuffle each dataset 20 times. 
Then we perform 20 independent trials for each of the methods on each dataset.
The average NMI results\footnote{We could not run DP-BMM(-FP) on the datasets SO-T and NTSO-T because of their longer running time that exceeds the capacity of the experimental computer.} of these runs are shown in Table~\ref{tbl_nmi_ho_vm}.

\begin{table}[H]
\captionsetup{font=small}
\centering
\caption{Normalized Mutual Information (NMI) score of different clustering methods. The highest result for a particular dataset is denoted bold.}
\label{tbl_nmi_ho_vm}
\renewcommand{\tabcolsep}{4pt}
\scriptsize
\begin{tabular}{|c|c|c|c|c|c|}
\hline
Clustering & Eva. & \multicolumn{4}{c|} { Data Sets} \\ \cline{3-6}
Methods & & Ns-T & Ts-T & SO-T & NTSO-T 
 \\ \cline{3-6}  \hline
FastStream-unigram & & 0.829 & 0.831 & 0.732   & 0.725  \\ \cline{3-6}
FastStream-bigram & & 0.837 & 0.826 & 0.754   & 0.729  \\ \cline{3-6}
FastStream-biterm & & 0.844 & 0.848 & 0.777   & 0.774   \\
\cline{1-1} \cline{3-6}
Fast-MStream & & 0.857 & 0.868 & 0.615 &  0.611    \\ \cline{3-6}
Fast-MStreamF & & 0.877  & 0.878 & 0.651  & 0.613  \\ \cline{3-6}
Fast-\citet{rakib:2020} & NMI &0.864 & 0.901 & 0.654 & 0.639  \\ 
\cline{1-1} \cline{3-6}
MStream & & 0.859 & 0.867 & 0.618 & 0.608 \\ \cline{3-6}
MStreamF &  & 0.879  & \textbf{0.877} &  0.652 & 0.619    \\  \cline{3-6}
DP-BMM & & \textbf{0.883}  & 0.862 & -- & --   \\ \cline{3-6} 
DP-BMM-FP & & 0.838 & 0.875 & --  & -- \\ \cline{3-6} 
OSDM &  & 0.858 & 0.842 & 0.463 & 0.441  \\ \cline{3-6}
\citet{rakib:2020} & & 0.862 & 0.892 & 0.658 & 0.641    \\
\hline

\end{tabular}
\end{table}


Our experimental results show that FastStream (-unigram,-bigram, and -biterm) perform significantly better than the state-of-the methods in terms of NMI on the dataset SO-T and the combined dataset NTSO-T.
Among the three variants of FastStream, FastStream-biterm performs better than FastStream-unigram and FastStream-bigram on all the datasets.
The reason is that, by using biterm features, we can extract sufficient contexts for short texts and the biterm features are more distinctive which in turn more likely to cluster the texts into their proper clusters~\citep{BMM2020}. 



The state-of-the-art methods MStream(F), DP-BMM(-FP), and OSDM perform better than FastStream on the datasets Ns-T and Ts-T since these methods tune the hyper parameters (e.g., $\alpha$, $\beta$) on these datasets to 
obtain optimal clustering performance.
On the contrary, our method uses dynamic similarity thresholds to assign texts to the clusters (new or existing) and 
does not require this kind of hyper parameter tuning on a particular dataset.

The overall performance of FastStream is comparable to the performance of the state-of-the-art methods on the datasets Ns-T and Ts-T.
In addition, our method significantly outperforms the state-of-the-art methods on the datasets SO-T and NTSO-T as the hyper parameters of the state-of-the-art methods were not tuned on these two datasets.

\subsubsection{Comparison of Running Time}
The average running times (in seconds) of FastStream and the state-of-the-art methods are shown in Table~\ref{tbl-runtime}.

\begin{table}[H]
\captionsetup{font=small}
\centering
\caption{Average running times (in seconds) of different methods.}
\label{tbl-runtime}
\renewcommand{\tabcolsep}{4pt}
\scriptsize
\begin{tabular}{|c|c|c|c|c|}
\hline
Clustering & \multicolumn{4}{c|} { Data Sets} \\ \cline{2-5}
Methods & Ns-T & Ts-T & SO-T & NTSO-T 
 \\ \cline{2-5}  \hline
FastStream-unigram & 3 & 20 & 85   & 168  \\ \cline{2-5}
FastStream-bigram & 3 & 18 & 86  & 167  \\ \cline{2-5}
FastStream-biterm & 4 & 24 & 101 & 183   \\ \cline{2-5}
 \hline
Fast-MStream  & 115 & 219 & 1318   & 1790  \\ \cline{2-5}
Fast-MStreamF & 83 & 193 & 811   & 1198  \\ \cline{2-5}
Fast-\citet{rakib:2020}  & 32 & 79 & 479   & 613 \\ 
 \hline
MStream & 227  & 541 & 3319 & 4289  \\ \cline{2-5}
MStreamF  & 173 & 295 & 1481 & 2137 \\  \cline{2-5}
DP-BMM &  5016 & 12307 & -- & --   \\ \cline{2-5} 
DP-BMM-FP & 880 & 2854 & --  & -- \\ \cline{2-5} 
OSDM  & 31 & 196 & 1103 & 1882  \\ \cline{2-5}
\citet{rakib:2020}  & 80 & 165 & 889  & 1389  \\
\hline

\end{tabular}
\end{table}

The running time of FastStream is several orders of magnitude faster than that of MStream(F), DP-BMM(-FP), and OSDM on all datasets.
In addition, we improve the running time of the existing methods (e.g., Fast-MStream, Fast-MStreamF, and Fast-\citet{rakib:2020}) by using inverted index~\citep{invertedindex:2014}.
The reason is that we do not compute similarity between a text and all clusters while assigning a text to a cluster.
Instead we select a specific set of clusters using the biterms of the text based on inverted index and compute similarities between the text and the selected clusters.
We store almost twice the number of biterms than other methods as we use inverted index to select a specific set of clusters for a particular text.
We consider this as a small price to pay for the significant improvement in running time of our proposed method (FastStream) and the existing state-of-the-art methods.




\section{Conclusion and Future Work}
We have demonstrated that building an efficient clustering model based on inverted index and by assigning texts to the clusters using dynamic similarity thresholds improves the clustering quality of our proposed method and outperforms the state-of-the-art short text stream clustering methods on larger datasets.

We also demonstrated that by computing similarity between a text and a specific set of clusters instead of all clusters, we significantly reduce the running time of our method (FastStream) and the faster versions of the state-of-the-art methods than that of the existing state-of-the-art methods. 
We contribute two new datasets SO-T and NTSO-T comprising of texts from various domains to examine how the FastStream and other state-of-the-art methods perform when texts from different domains arrive in random order. 

In the future, we plan to cluster significantly larger short text streams (composed of a few million texts). 
We also plan to apply our improved text stream clustering algorithm to better identify
groups of users of social media platforms interested in similar topics.

\bibliographystyle{ACM-Reference-Format}
\bibliography{sample-base}


\begin{thebibliography}{14}


\ifx \showCODEN    \undefined \def \showCODEN     #1{\unskip}     \fi
\ifx \showDOI      \undefined \def \showDOI       #1{#1}\fi
\ifx \showISBNx    \undefined \def \showISBNx     #1{\unskip}     \fi
\ifx \showISBNxiii \undefined \def \showISBNxiii  #1{\unskip}     \fi
\ifx \showISSN     \undefined \def \showISSN      #1{\unskip}     \fi
\ifx \showLCCN     \undefined \def \showLCCN      #1{\unskip}     \fi
\ifx \shownote     \undefined \def \shownote      #1{#1}          \fi
\ifx \showarticletitle \undefined \def \showarticletitle #1{#1}   \fi
\ifx \showURL      \undefined \def \showURL       {\relax}        \fi
\providecommand\bibfield[2]{#2}
\providecommand\bibinfo[2]{#2}
\providecommand\natexlab[1]{#1}
\providecommand\showeprint[2][]{arXiv:#2}

\bibitem[\protect\citeauthoryear{Aggarwal, Han, Wang, and Yu}{Aggarwal
  et~al\mbox{.}}{2003}]%
        {clustream:2003}
\bibfield{author}{\bibinfo{person}{Charu~C. Aggarwal}, \bibinfo{person}{Jiawei
  Han}, \bibinfo{person}{Jianyong Wang}, {and} \bibinfo{person}{Philip~S. Yu}.}
  \bibinfo{year}{2003}\natexlab{}.
\newblock \showarticletitle{A Framework for Clustering Evolving Data Streams}.
  In \bibinfo{booktitle}{\emph{Proceedings of the 29th International Conference
  on Very Large Data Bases}} (Berlin, Germany). \bibinfo{pages}{81--92}.
\newblock


\bibitem[\protect\citeauthoryear{Blei and Lafferty}{Blei and Lafferty}{2006}]%
        {DTM:2006}
\bibfield{author}{\bibinfo{person}{David~M. Blei} {and}
  \bibinfo{person}{John~D. Lafferty}.} \bibinfo{year}{2006}\natexlab{}.
\newblock \showarticletitle{Dynamic Topic Models}. In
  \bibinfo{booktitle}{\emph{Proceedings of the 23rd International Conference on
  Machine Learning}} (Pittsburgh, Pennsylvania, USA). \bibinfo{publisher}{ACM},
  \bibinfo{address}{New York, NY, USA}, \bibinfo{pages}{113--120}.
\newblock


\bibitem[\protect\citeauthoryear{Carnein and Trautmann}{Carnein and
  Trautmann}{2019}]%
        {DataSurvey2019}
\bibfield{author}{\bibinfo{person}{Matthias Carnein} {and}
  \bibinfo{person}{Heike Trautmann}.} \bibinfo{year}{2019}\natexlab{}.
\newblock \showarticletitle{Optimizing Data Stream Representation: An Extensive
  Survey on Stream Clustering Algorithms}.
\newblock \bibinfo{journal}{\emph{Business {\&} Information Systems
  Engineering}} \bibinfo{volume}{61}, \bibinfo{number}{3} (\bibinfo{date}{01
  Jun} \bibinfo{year}{2019}), \bibinfo{pages}{277--297}.
\newblock


\bibitem[\protect\citeauthoryear{Chen, Gong, and Liu}{Chen
  et~al\mbox{.}}{2020}]%
        {BMM2020}
\bibfield{author}{\bibinfo{person}{Junyang Chen}, \bibinfo{person}{Zhiguo
  Gong}, {and} \bibinfo{person}{Weiwen Liu}.} \bibinfo{year}{2020}\natexlab{}.
\newblock \showarticletitle{A Dirichlet process biterm-based mixture model for
  short text stream clustering}.
\newblock \bibinfo{journal}{\emph{Applied Intelligence}} \bibinfo{volume}{50},
  \bibinfo{number}{5} (\bibinfo{year}{2020}), \bibinfo{pages}{1609--1619}.
\newblock


\bibitem[\protect\citeauthoryear{Erk}{Erk}{2012}]%
        {VSM2012}
\bibfield{author}{\bibinfo{person}{Katrin Erk}.}
  \bibinfo{year}{2012}\natexlab{}.
\newblock \showarticletitle{Vector Space Models of Word Meaning and Phrase
  Meaning: A Survey}.
\newblock \bibinfo{journal}{\emph{Language and Linguistics Compass}}
  \bibinfo{volume}{6} (\bibinfo{year}{2012}), \bibinfo{pages}{635--653}.
\newblock


\bibitem[\protect\citeauthoryear{{Ilic}, {Spalevic}, and {Veinovic}}{{Ilic}
  et~al\mbox{.}}{2014}]%
        {invertedindex:2014}
\bibfield{author}{\bibinfo{person}{M. {Ilic}}, \bibinfo{person}{P. {Spalevic}},
  {and} \bibinfo{person}{M. {Veinovic}}.} \bibinfo{year}{2014}\natexlab{}.
\newblock \showarticletitle{Inverted index search in data mining}. In
  \bibinfo{booktitle}{\emph{2014 22nd Telecommunications Forum Telfor
  (TELFOR)}}. \bibinfo{pages}{943--946}.
\newblock


\bibitem[\protect\citeauthoryear{Ishwaran and James}{Ishwaran and
  James}{2001}]%
        {gibssampling:2001}
\bibfield{author}{\bibinfo{person}{Hemant Ishwaran} {and}
  \bibinfo{person}{{Lancelot F.} James}.} \bibinfo{year}{2001}\natexlab{}.
\newblock \showarticletitle{Gibbs Sampling Methods for Stick-Breaking Priors}.
\newblock \bibinfo{journal}{\emph{J. Amer. Statist. Assoc.}}
  \bibinfo{volume}{96}, \bibinfo{number}{453} (\bibinfo{date}{1 3}
  \bibinfo{year}{2001}), \bibinfo{pages}{161--173}.
\newblock
\showISSN{0162-1459}


\bibitem[\protect\citeauthoryear{Kalogeratos, Zagorisios, and
  Likas}{Kalogeratos et~al\mbox{.}}{2016}]%
        {textstream2016}
\bibfield{author}{\bibinfo{person}{Argyris Kalogeratos},
  \bibinfo{person}{Panagiotis Zagorisios}, {and} \bibinfo{person}{Aristidis
  Likas}.} \bibinfo{year}{2016}\natexlab{}.
\newblock \showarticletitle{Improving Text Stream Clustering Using Term
  Burstiness and Co-Burstiness}. In \bibinfo{booktitle}{\emph{Proceedings of
  the 9th Hellenic Conference on Artificial Intelligence}} (Thessaloniki,
  Greece) \emph{(\bibinfo{series}{SETN '16})}. \bibinfo{publisher}{Association
  for Computing Machinery}, \bibinfo{address}{New York, NY, USA}, Article
  \bibinfo{articleno}{16}, \bibinfo{numpages}{9}~pages.
\newblock
\showISBNx{9781450337342}


\bibitem[\protect\citeauthoryear{Kumar, Shao, Uddin, and Ali}{Kumar
  et~al\mbox{.}}{2020}]%
        {osdm:2020}
\bibfield{author}{\bibinfo{person}{Jay Kumar}, \bibinfo{person}{Junming Shao},
  \bibinfo{person}{Salah Uddin}, {and} \bibinfo{person}{Wazir Ali}.}
  \bibinfo{year}{2020}\natexlab{}.
\newblock \showarticletitle{An Online Semantic-enhanced {D}irichlet Model for
  Short Text Stream Clustering}. In \bibinfo{booktitle}{\emph{Proc. of the 58th
  Annual Meeting of the Association for Computational Linguistics}}.
  \bibinfo{address}{Online}, \bibinfo{pages}{766--776}.
\newblock


\bibitem[\protect\citeauthoryear{Liang, Yilmaz, and Kanoulas}{Liang
  et~al\mbox{.}}{2016}]%
        {DCT-L:2016}
\bibfield{author}{\bibinfo{person}{Shangsong Liang}, \bibinfo{person}{Emine
  Yilmaz}, {and} \bibinfo{person}{Evangelos Kanoulas}.}
  \bibinfo{year}{2016}\natexlab{}.
\newblock \showarticletitle{Dynamic Clustering of Streaming Short Documents}.
  In \bibinfo{booktitle}{\emph{Proceedings of the 22nd ACM SIGKDD International
  Conference on Knowledge Discovery and Data Mining}} (San Francisco,
  California, USA). \bibinfo{pages}{995--1004}.
\newblock


\bibitem[\protect\citeauthoryear{Rakib, Zeh, and Milios}{Rakib
  et~al\mbox{.}}{2020}]%
        {rakib:2020}
\bibfield{author}{\bibinfo{person}{Md~Rashadul~Hasan Rakib},
  \bibinfo{person}{Norbert Zeh}, {and} \bibinfo{person}{Evangelos Milios}.}
  \bibinfo{year}{2020}\natexlab{}.
\newblock \showarticletitle{Short Text Stream Clustering via Frequent Word
  Pairs and Reassignment of Outliers to Clusters}
  \emph{(\bibinfo{series}{DocEng '20})}. \bibinfo{publisher}{Association for
  Computing Machinery}, \bibinfo{address}{New York, NY, USA}, Article
  \bibinfo{articleno}{13}, \bibinfo{numpages}{4}~pages.
\newblock
\showISBNx{9781450380003}


\bibitem[\protect\citeauthoryear{Shou, Wang, Chen, and Chen}{Shou
  et~al\mbox{.}}{2013}]%
        {Sumblr:2013}
\bibfield{author}{\bibinfo{person}{Lidan Shou}, \bibinfo{person}{Zhenhua Wang},
  \bibinfo{person}{Ke Chen}, {and} \bibinfo{person}{Gang Chen}.}
  \bibinfo{year}{2013}\natexlab{}.
\newblock \showarticletitle{Sumblr: Continuous Summarization of Evolving Tweet
  Streams}. In \bibinfo{booktitle}{\emph{Proceedings of the 36th International
  ACM SIGIR Conference on Information Retrieval}} (Dublin, Ireland).
  \bibinfo{pages}{533--542}.
\newblock


\bibitem[\protect\citeauthoryear{Yin, Chao, Liu, Zhang, Yu, and Wang}{Yin
  et~al\mbox{.}}{2018}]%
        {Mstream:2018}
\bibfield{author}{\bibinfo{person}{Jianhua Yin}, \bibinfo{person}{Daren Chao},
  \bibinfo{person}{Zhongkun Liu}, \bibinfo{person}{Wei Zhang},
  \bibinfo{person}{Xiaohui Yu}, {and} \bibinfo{person}{Jianyong Wang}.}
  \bibinfo{year}{2018}\natexlab{}.
\newblock \showarticletitle{Model-based Clustering of Short Text Streams}. In
  \bibinfo{booktitle}{\emph{Proceedings of the 24th ACM SIGKDD International
  Conference on Knowledge Discovery and Data Mining}} (London, United Kingdom).
  \bibinfo{pages}{2634--2642}.
\newblock


\bibitem[\protect\citeauthoryear{Zhao, Liang, Ren, Ma, Yilmaz, and
  de~Rijke}{Zhao et~al\mbox{.}}{2016}]%
        {usercluster:2016}
\bibfield{author}{\bibinfo{person}{Yukun Zhao}, \bibinfo{person}{Shangsong
  Liang}, \bibinfo{person}{Zhaochun Ren}, \bibinfo{person}{Jun Ma},
  \bibinfo{person}{Emine Yilmaz}, {and} \bibinfo{person}{Maarten de Rijke}.}
  \bibinfo{year}{2016}\natexlab{}.
\newblock \showarticletitle{Explainable User Clustering in Short Text Streams}.
  In \bibinfo{booktitle}{\emph{Proceedings of the 39th International ACM SIGIR
  Conference on Research and Development in Information Retrieval}} (Pisa,
  Italy). \bibinfo{publisher}{ACM}, \bibinfo{pages}{155--164}.
\newblock
\showISBNx{978-1-4503-4069-4}


\end{thebibliography}

\end{document}